\documentclass[twocolumn]{revtex4}
\usepackage{amsmath, amsfonts, amssymb, graphicx, color}

\definecolor{cerisepink}{rgb}{0.93, 0.23, 0.51}

\newcommand{\rev}{}

\newcommand{\lastequal}{Corresponding authors. These authors contributed equally.}

\begin{document}

\newcommand{\deftitle}{{Learning the statistics and landscape of somatic mutation-induced insertions and deletions in antibodies}}

\title{\deftitle}

\author{Cosimo Lupo}
\thanks{Current address: Istituto Nazionale di Fisica Nucleare (INFN), Sezione di Roma I, 00185 Rome, Italy.}
\affiliation{Laboratoire de physique de l'\'Ecole normale sup\'erieure,
  CNRS, PSL University, Sorbonne Universit\'e, and Universit\'e de
  Paris, 75005 Paris, France}
\author{Natanael Spisak}
\affiliation{Laboratoire de physique de l'\'Ecole normale sup\'erieure,
  CNRS, PSL University, Sorbonne Universit\'e, and Universit\'e de
  Paris, 75005 Paris, France}
\author{Aleksandra M. Walczak}
\thanks{\lastequal}
\affiliation{Laboratoire de physique de l'\'Ecole normale sup\'erieure,
  CNRS, PSL University, Sorbonne Universit\'e, and Universit\'e de
  Paris, 75005 Paris, France}
\author{Thierry Mora}
\thanks{\lastequal}
\affiliation{Laboratoire de physique de l'\'Ecole normale sup\'erieure,
  CNRS, PSL University, Sorbonne Universit\'e, and Universit\'e de
  Paris, 75005 Paris, France}

\begin{abstract}
Affinity maturation is crucial for improving the binding affinity of antibodies to antigens. This process is mainly driven by point substitutions caused by somatic hypermutations of the immunoglobulin gene. It also includes deletions and insertions of genomic material known as indels. While the landscape of point substitutions has been extensively studied, a detailed statistical description of indels is still lacking. Here we present a probabilistic inference tool to learn the statistics of indels from repertoire sequencing data, which overcomes the pitfalls and biases of standard annotation methods. The model includes antibody-specific maturation ages to account for variable mutational loads in the repertoire. After validation on synthetic data, we applied our tool to a large dataset of human immunoglobulin heavy chains. The inferred model allows us to identify universal statistical features of indels {\rev in heavy chains}. We report distinct insertion and deletion hotspots, and show that the distribution of lengths of indels follows a geometric distribution, which puts constraints on future mechanistic models of the hypermutation process.

\end{abstract}

\maketitle

\section{Introduction}
The extraordinary ability of the immune system to detect and neutralize pathogens is partly assured by B cells, with the production of huge amounts of diverse immunoglobulins (Ig). Their initial diversity, provided by  stochastic V(D)J recombination~\cite{HozumiTonegawa1976, BoydEtAl2009, GlanvilleEtAl2009, LarimoreEtAl2012, ElhanatiEtAl2015, DeWittEtAl2016, MarcouEtAl2018, BrineyEtAl2019}, is further increased by affinity maturation, a mutation-selection Darwinian process that takes place {\rev primarily} in germinal centers {\rev (as well as extrafollicularly prior to germinal center formation~\cite{ElsnerShlomchik2020})}, where cells with the highest affinity to the pathogen compete for survival and proliferation~\cite{VictoraNussenzweig2012, CobeyEtAl2015, MesinEtAl2016}.

During affinity maturation, cells diversify their immunoglobulin receptors through somatic hypermutations (SHM), which are initiated by the activation-induced cytidine deaminase (AID) enzyme~\cite{FengEtAl2020} targeting immunoglobulin-coding genes. Its main effect is to cause point substitutions with a much higher rate than classical somatic mutations, namely $\sim 10^{-3}$ substitutions per base-pair per cell division~\cite{KleinsteinEtAl2003, OdegardSchatz2006}. Since affinity maturation implies many proliferation and selection rounds, this process can lead to Ig with up to 30-40\,\% of amino-acid substitutions with respect to the initial V(D)J rearrangement.
Statistics of SHM have been extensively studied~\cite{YaariEtAl2013, McCoyEtAl2015, ElhanatiEtAl2015, CuiEtAl2016, ShengEtAl2017, HoehnEtAl2017, MarcouEtAl2018, DharEtAl2018, SpisakEtAl2020}, mainly thanks to high-throughput repertoire sequencing~\cite{BoydEtAl2009, GlanvilleEtAl2009, DeWittEtAl2016, BrineyEtAl2019}, although the link to the molecular details of AID functioning remains elusive. 

But point mutations are not the only modifications found in matured Ig receptors. Deletions and insertions of one or multiple base pairs, collectively referred to as \textit{indels}, have been detected in the V and J segments, albeit at a much lower rate than substitutions.
Indels have been reported both in both heavy and light Ig chains~\cite{WilsonEtAl1998a, WilsonEtAl1998b, KleinEtAl1998, FischerKuppers1998, GoossensEtAl1998, OhlinBorrebaeck1998, deWildtEtAl1999b, KuppersEtAl1999, BemarkNeuberger2003, ReasonZhou2006}, with a frequency that grows with the substitution rate during affinity maturation~\cite{WilsonEtAl1998a, WilsonEtAl1998b, deWildtEtAl1999b, OhlinBorrebaeck1998}. Indels occur non-uniformly along the V germline templates, with a preference towards complementary-determining regions (CDRs) with respect to framework regions (FWRs); these `hotspots' often coincide with those for point mutations~\cite{WilsonEtAl1998b, ReasonZhou2006, BrineyEtAl2012, YeapEtAl2015}.

Indels are known to be critical for the increase of antibody repertoire diversity, enhancing the immune response in the presence of viral and bacterial pathogens as e.\,g. HIV-1 and influenza~\cite{ZhouEtAl2004, WuEtAl2010, WalkerEtAl2009, WalkerEtAl2011, KeplerEtAl2014} and in response to vaccination~\cite{ReasonZhou2006}. Several works have shown a beneficial impact of indels for binding affinity and neutralization activity \textit{in vitro} and \textit{in vivo}~\cite{ReasonZhou2006, KrauseEtAl2011, PejchalEtAl2011}, opening new evolutionary pathways beyond point mutations. A particularly striking example is given by anti-HIV-1 broadly-neutralizing antibodies (bnAbs), many of which have a high indel load~\cite{WuEtAl2011, MascolaHaynes2013, SteichenEtAl2019}. In particular, bnAb CH31 was shown to owe its neutralization breadth to a particular long insertion in its lineage~\cite{KeplerEtAl2014}.

Yet, little is known about the biological mechanisms behind the occurence of indels. Observations show that inserted segments have a high homology with their flanking regions (either on the $3'$ or $5'$ sides)~\cite{deWildtEtAl1999b, BemarkNeuberger2003, ReasonZhou2006, BrineyEtAl2012}, possibly suggesting duplications rather than non-templated, random insertions. A proposed mechanism~\cite{StreisingerEtAl1966, GoldingEtAl1987} involves template regions with repeats and palindromic sequences, where the polymerase can ``slip'' during replication and ``jump'' to a nearby homologous element on the templated strand, producing a loop in either the daughter or the templated strand. If this unpaired loop is not properly repaired by the specific DNA repair mechanisms, according to its location on the daughter or the templated strand, it will be propagated as an insertion or a deletion, respectively. This mechanism would also explain co-localization of indels and SHM, since they could likely occur independently in the same susceptible regions (hot-spots due to repeats and palindromes) or also as a product of the repair mechanisms {\rev targeting} existing replication errors.

Deep high-throughput sequencing of Ig repertoires now allows us to gain insight into the mechanisms of indels, and to collect statistics about their {\rev frequency}, length distributions, and positional preference. By contrast, previous studies \cite{BrineyEtAl2012, YeapEtAl2015} have relied on relatively small datasets and therefore small number of indel events due to their rarity. In addition, most studies have focused on productive sequences, for which it is difficult to decouple intrinsic preferences from selection effects. Here, we exploit a recently published large immunoglobulin heavy chain (IgH) repertoire dataset comprising sequences from 9 healthy donors~\cite{BrineyEtAl2019}. We {\rev developed} an inference methodology to provide robust indel statistics and to capture their universal features in the face of natural individual variability.

Our approach uses a probabilistic framework to overcome the issue of annotation errors. Classically, indels are identified by looking at the best-scoring annotation. Such deterministic annotation is sometimes wrong, and we will show that this leads to systematic biases. Our probabilistic method, which considers multiple indel and {\rev point-mutation} scenarios weighted by their probabilities, allows us to remove these biases, in a similar way to what was done for inferring the statistics of VDJ recombination~\cite{MuruganEtAl2012,ElhanatiEtAl2015,MarcouEtAl2018}. A second key ingredient of our approach is to assume that each sequence may have a different mutation rate, since repertoire data combines B cells from various stages of affinity maturation. We first tested and validated our method on synthetically produced data. We then applied it to Ig sequences with a frameshift in their CDR3 region, which are believed to be nonproductive and thus free of selection biases. This allows us to characterize in great detail the statistics and intrinsic preferences of indel generation.

\section{Results}
\subsection*{Insertion and deletion events scale with the mutation rate}
We conduct our analysis on a recently published high-throughput IgH repertoire dataset, obtained from the blood of 9 healthy donors~\cite{BrineyEtAl2019}. IgM and IgG expressing cells were isolated and analyzed separately. Using raw sequence data, we further segregated IgH sequences into productive (P) and nonproductive (NP) sequences, depending on whether their CDR3 had a frameshift or not. Since hypermutation indels are rare (see below) compared to VDJ recombination frameshifts, we assumed that nonproductive sequences were already faulty at generation. Cells with NP sequences owe their survival to a successful IgH rearrangement on the second chromosome, meaning that the NP sequences themselves are free of selection effects.
 Most productive sequences undergo various selection processes, which bias the statistics of their features~\cite{ElhanatiEtAl2015}.
Here we mainly restrict our analysis to nonproductive sequences to capture the biochemistry of the hypermutation machinery.
{\rev We obtained $421,185$ IgG and $459,165$ IgM nonproductive sequences from 9 donors.}

\begin{figure*}
\begin{center}
\includegraphics[width=\textwidth]{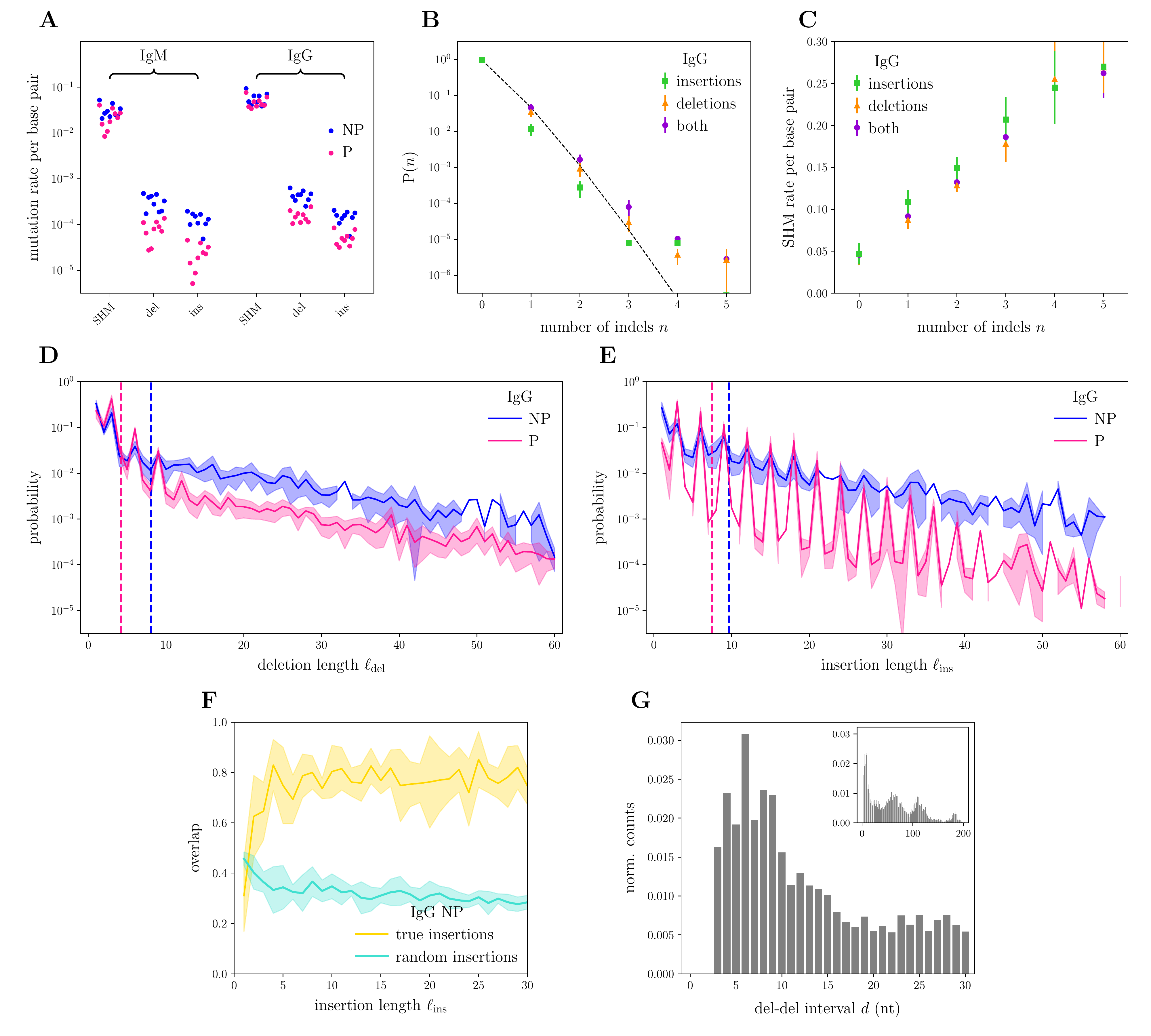}
\caption{\textbf{{\rev Indel statistics} in real IgG sequences from deterministic annotation}. \textbf{(A)} Somatic hyper-mutations (SHM), deletion (del) and insertion (ins) absolute rates per base pair for the 9 Briney donors~\cite{BrineyEtAl2019}. Segregation into productive (P) and nonproductive (NP) sequences coded by colors; data points from IgM sequences are also reported, for comparison. \textbf{(B)} Fraction of IgG sequences (productive and nonproductive) with exactly $n$ indels, either separated by type (del or ins) or cumulated; mean and standard deviation over the 9 donors represented by the markers and bars, respectively. The dashed black line represents the closest Poisson trend for cumulated indels of both types. \textbf{(C)} Fraction of SHM (in nucleotides) along IgG sequences (productive and nonproductive) conditioned on having exactly $n$ indels, again either cumulated or separated by type; averages and one standard deviation bars over the 9 donors. \textbf{(D)} Length profiles for deletions, after NP-P segregation. Average and standard deviation over the 9 donors represented by the solid line and the shaded area around, respectively; corresponding average lengths are represented by the two vertical dashed lines. \textbf{(E)} Same as (D), but for insertions. \textbf{(F)} Overlap between inserted base pairs and same-length flanking regions on either $3'$ or $5'$ side along the sequence (larger is considered); average and one standard deviation over the 9 donors given by solid line and shaded area around, respectively. For comparison, overlap between random insertions and flanking regions is also reported (details in the main text). \textbf{(G)} Distribution of distances between deletions along the same sequence, {\rev measured as the number of base pairs in between}, for IgG sequences (productive and nonproductive) hosting exactly two deletions and no insertions (data pooled together from the 9 donors); inset shows the full distribution, main plot focuses on the short-distance region.}
\label{fig:real_data_general_stats}
\end{center}
\end{figure*}

To get preliminary insights into indel statistics, we first annotated each sequence deterministically using {\rev IgBLAST}~\cite{YeEtAl2013}.
We estimated the rates per base pair of SHM point mutations, deletions and insertions, in both IgM and IgG compartments, separately for each individual and each of the productive and nonproductive subsets (Fig.~\ref{fig:real_data_general_stats}A). Mutation rates for productive sequences are consistent with previous estimates~\cite{BrineyEtAl2012}, with more mutations in IgG than in IgM. Some individual variability is present, suggesting variations in the degree of antigen exposure across individuals. Nonproductive sequences have higher rates of SHM than productive ones, especially for insertions and deletions ($\sim$10-fold difference for IgM and a bit less for IgG), suggesting that mutations are mainly deleterious.

Fig.~\ref{fig:real_data_general_stats}B shows the distribution of the number of indels in the IgG subset across donors. For comparison, a Poisson-distribution fit agrees well with the data, but underestimates sequences with many indels.
Point mutations and indels both accumulate during affinity maturation and are mediated by AID~\cite{WilsonEtAl1998b, OhlinBorrebaeck1998, deWildtEtAl1999b, ReasonZhou2006, BrineyEtAl2012, YeapEtAl2015}. Indels are mostly present in highly point-mutated sequences. Both types of mutations have been reported to co-localize along the sequence, preferably in the CDRs~{\rev \cite{BrineyEtAl2012, YeapEtAl2015}}. They have also been shown to co-localize temporally, as deduced from the analysis of antibody phylogenetic trees~\cite{ReasonZhou2006, KeplerEtAl2014}. Consistent with this picture, we observe that the average point-mutation rate {\rev per base pair} in IgG grows linearly with the number of indel events (Fig.~\ref{fig:real_data_general_stats}C).

\subsection*{Length and composition of inserted and deleted segments}

Deletion and insertion length profiles, which show the distributions of the number of deleted or inserted base pairs, are shown in Figs.~\ref{fig:real_data_general_stats}D and E. Signatures of selection can be seen through an increased preference for multiples of 3 in productive {\em versus} nonproductive sequences, suggesting that indels that induce a frameshit are deleterious.
The effect is only present at short deletion lengths, but holds for {\rev long insertion lengths}. Note that a weaker 3-fold periodicity is also present in nonproductive sequences, despite them being in principle free of selection effects. This could come from sequences that were previously productive but {\rev were  frameshifted} relatively recently due to an indel in the CDR3.
Long insertions are favored in productive sequences {\rev in comparison to} long deletions, probably because of stability and functionality constraints \cite{BrineyEtAl2012}. By constrast, in nonproductive sequences deletions and insertions follow similar statistics, suggesting a common mechanism.
Both insertions and deletions are approximately geometrically distributed, at least in the tail that describes large inserted or deleted segments. This geometric law puts strong constraints on the type of biophysical mechanisms by which indels may be created.

A proposed mechanism for indels is the \textit{polymerase slippage model}~\cite{StreisingerEtAl1966, GoldingEtAl1987}. In germline regions that have repeats and palindromic sequences, the polymerase can ``slip'' during replication and ``jump'' to the closeby homologous element on the templated strand, resulting in a insertion or duplication depending on the location of the polymerase on the daughter or templated strand. In this scenario, insertions would be the result of duplications, and would thus be expected to be homologous to their flanking regions. The analysis of inserted segments in nonproductive sequence does reveal a higher than expected overlap with the best matching of the two immediately flanking sequences (Fig.~\ref{fig:real_data_general_stats}F). Previous similar observations were made on small numbers of productive sequences \cite{BrineyEtAl2012}, for which effects of selection may confound the effect.
Note that overlap is substantially different from 1 ($\approx 0.8$), suggesting $20\%$ additional errors in the duplication, which is much higher than the mean point-substitution rate {\rev per base pair} (around $5-6\%$). Also, we find a weak preference for the $5'$ end as a duplication source~(Fig.~\ref{fig:SI_overlap_ins}).
However, we found no significant overlap between deleted segments and their flanking regions~(Fig.~\ref{fig:SI_overlap_del}). {\rev This suggests either that the slip-and-jump mechanism does not favor homologous regions for deletions, or that other mechanisms are responsible for deletions, such as double-strand breaks \cite{YeapEtAl2015,HwangEtAl2017}}.

\begin{figure*}
\begin{center}
\includegraphics[width=\textwidth]{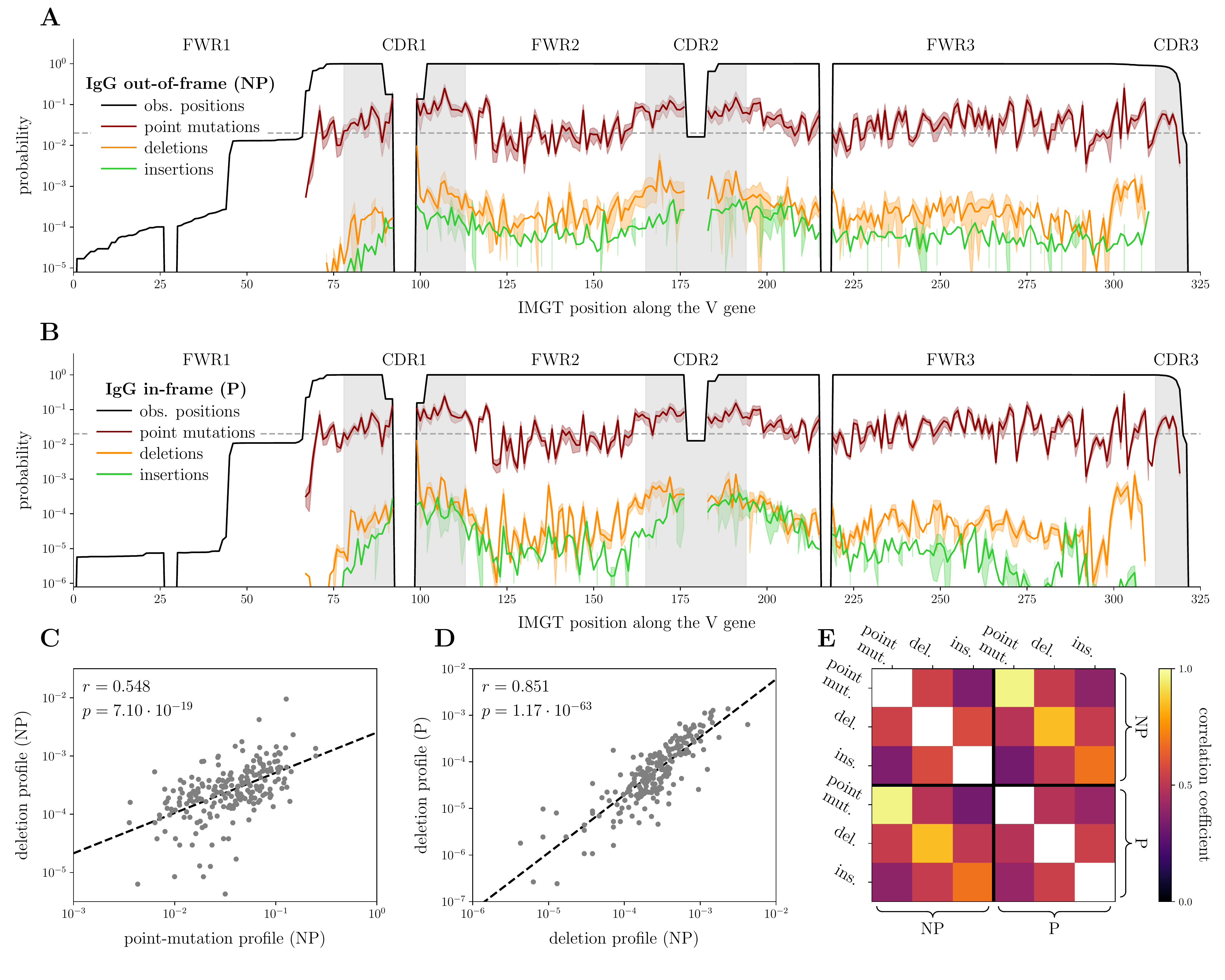}
\caption{\textbf{{\rev Indel profiles}}. \textbf{(A)} Mutation profiles for out-of-frame (NP) sequences, pooling together all the 9 Briney donors~\cite{BrineyEtAl2019} and all the V templates {\rev (with shaded area representing individual variability)}, using the absolute IMGT indexing of positions~\cite{GiudicelliEtAl2006}. Due to alignment gaps and finite read lengths, many IMGT positions are only observed in a fraction of sequences (black line). Mutation profiles are rescaled to account for this, and are masked for IMGT positions appearing in less than $2\%$ of sequences (dashed horizontal line). Indel rates past IMGT position 310 are also not reported due to ambiguities with junctional diversity. \textbf{(B)} Same as (A), for in-frame (P) sequences. \textbf{(C)} Scatter plot between the SHM profile and the deletion profile for NP sequences; linear regression line is reported, together with numerical results (correlation coefficient $r$ and $p$ value). \textbf{(D)} Same as (C), here comparing deletion profiles for NP sequences versus P sequences. \textbf{(E)} Correlation between all the mutation profiles, computed either on NP or on P sequences.}
\label{fig:real_data_V_profiles}
\end{center}
\end{figure*}

\subsection*{Indels co-localize with each other and with point mutations even in absence of selection}
Since point mutations tend to co-localize in hotspot regions, and indels are {\rev associated with} point mutations, we wondered if indels co-localized as well.
Fig.~\ref{fig:real_data_general_stats}G shows the distribution of distances between two consecutive deletion events along the same sequence. To rule out possible confounding factors, we focused on sequences with exactly two deletion events and no insertions. Since such sequences are rare, we pooled the IgG repertoires (P and NP) of all individuals for this analysis. The full distribution (inset) is highly non-uniform, consistent with the observation that indels concentrate on CDRs~\cite{BrineyEtAl2012, YeapEtAl2015}. In addition, we observe a striking depletion of deletion pairs separated by less than 3 base pairs. This depletion is an artifact of the alignment software, which merges deletion events if they are too close to each other. This effect implies that the indels involving $\approx 50-60$ base pairs reported in Figs.~\ref{fig:real_data_general_stats}D-E may result from such mergers, which standard alignment tools cannot disentangle. Our probabilistic framework will allow us to address this issue in the following sections.

We then studied the positional preference of indels along the V segment. To establish a universal profile across all V genes, we used the gapped absolute indexing provided by IMGT~\cite{GiudicelliEtAl2006}, and pooled the sequences of all 9 donors.
The resulting profiles for nonproductive and productive sequences are shown in Figs.~\ref{fig:real_data_V_profiles}A and B.
Due to alignment gaps and finite read lengths, many IMGT positions are only observed in a fraction of sequences (black line). 
The indel profiles are highly non-uniform, with generally higher rates in the CDRs than in the Framework Regions (FWR), with the exception of a deletion hotspot the end of FWR3. We note that these trends, which are consistent with previous reports on productive sequences~\cite{BrineyEtAl2012, YeapEtAl2015}, are found also in nonproductive sequences, suggesting that they cannot be attributed to selective or functional effects.

The different mutation profiles (point mutations, insertions, deletions, productive versus nonproductive) are correlated (Figs.~\ref{fig:real_data_V_profiles}C-E), meaning that indel and point-mutation hotspots are largely shared. Deletion profiles are well correlated with both point-mutation and insertion profiles. {\rev These correlations are larger than between insertions and point mutations, suggesting a possibly different mechanism for the two types of indels. In addition and consistently with this picture, deletions and point mutations show an enrichment of co-localization (within a few base pairs) relative to what would be expected from their shared positional preferences, while insertions and point mutations do not (Fig.~\ref{fig:SI_corr_SHM_del}). These observations are consistent with the leading model that deletions are caused by double-strand breaks~\cite{YeapEtAl2015,HwangEtAl2017}.}

Profiles in productive versus nonproductive sets are very similar, suggesting that the {\rev positional} effects of functional selection on indels and point-mutation preferences is weak, {\rev and that most of the previously reported biases in indel positions \cite{BrineyEtAl2012} are due to the positional preference of the involved enzymes rather than to selection.}

\subsection*{A probabilistic alignment algorithm}

\begin{figure*}
\begin{center}
\includegraphics[width=\textwidth]{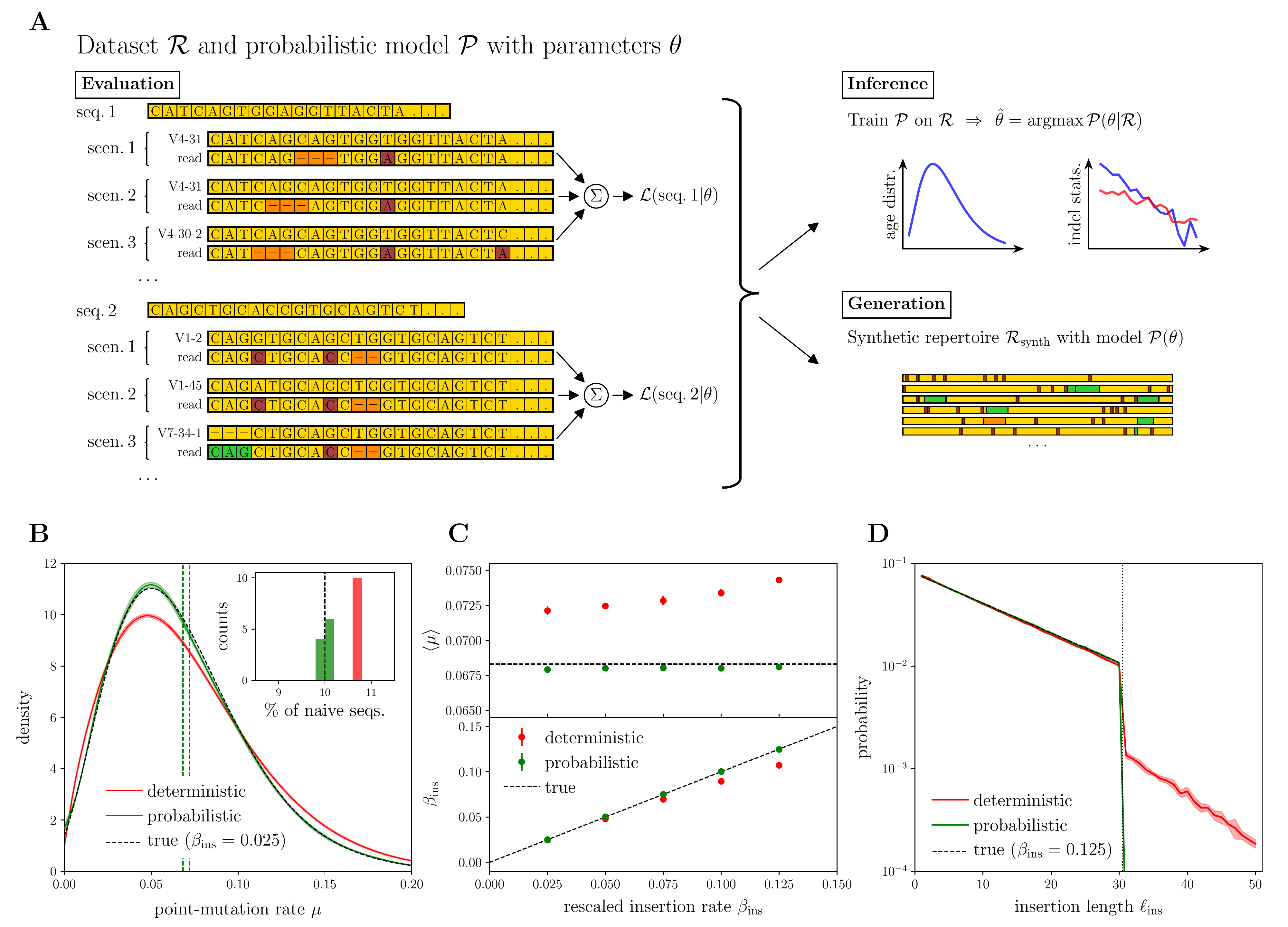}
\caption{\textbf{Workflow of the probabilistic framework and validation on synthetic data}. \textbf{(A)} Given the set of parameters~$\theta$ for the probabilistic model $\mathcal{P}$, for each sequence a set of alignment scenarios (given by template, read and SHM-indel pattern) is considered and the cumulated alignment likelihood $\mathcal{L}(\mathrm{seq.}|\theta)$ from all of them is computed. {\rev Indel statistics} and repertoire-wide age distribution from $\mathcal{R}$ can be extracted by using a given model (\textit{evaluation} mode) or by training a new model on $\mathcal{R}$ itself (\textit{inference} mode). Finally, a synthetic repertoire of arbitrary size can be generated, according to a given input model $\mathcal{P}(\theta)$ (\textit{generation} mode). \textbf{(B)} Repertoire-wide point-mutation rate distribution $P(\mu)$; mean and standard deviation over $\mathcal{N}=10$ independent realizations of a $100,000$ synthetic sequences repertoire represented as a solid line with the shaded area around, for both deterministic and probabilistic estimates. In the inset, the $\mathcal{N}$ estimations of the fraction of truly {\rev naive} sequences; color coding is the same as the main plot. \textbf{(C)} Performance of the probabilistic alignment software vs deterministic one when increasing the density of indels; estimation of the average point-mutation rate in the upper panel, and of {\rev the ratio of insertion to point-mutation rates $\beta_{\rm ins}$ in the lower panel} (the deletion rate {\rev $\beta_{\rm del}$} has a very similar behavior, shown in Fig.~\ref{fig:SI_synth_del}). {\rev Average values} plus one standard deviation error bars are obtained over the $\mathcal{N}$ independent synthetic repertoires. \textbf{(D)} Insertion length profiles for the largest value of $\beta_{\mathrm{ins}}$ considered in panel (C); mean and one standard deviation over the $\mathcal{N}$ independent realizations shown. Deletion length profile is again completely analogous, shown in Fig.~\ref{fig:SI_synth_del}. The vertical dotted line signals the maximal size $\ell=30$ used for generating insertions and deletions, then correctly identified by the probabilistic approach.}
\label{fig:synth_data_inference}
\end{center}
\end{figure*}

While deterministic annotations give us a good picture of indel statistics, they may introduce systematic biases that could confound the analysis, as e.g. the spurious gap in the inter-deletion interval distribution  (Fig.~\ref{fig:real_data_general_stats}G), or the over-estimation of long indel events. {\rev To give just an example of possible mistakes by deterministic annotation algorithms, a long indel observed in a sequence may have in fact been caused by several shorter indel events at the same site or close by. To evaluate the probability of the end product, one should consider all possible such decompositions into two or more indel events, and sum their probabilities.}
To address this issue, we designed a probabilistic approach to annotate and analyze hypermutated V segments, similarly to what was previously proposed to infer the statistics of V(D)J recombination from B- and T-cell receptor repertoires~\cite{MuruganEtAl2012,ElhanatiEtAl2015} and implemented in the IGoR tool \cite{MarcouEtAl2018}.

We first define a generative model of point mutations, insertions and deletions, with unknown parameters that describe the main statistics of the mutation process: distribution of SHM rates across sequences, mean ratios of deletion and insertion rates to the point mutation rate, and insertion and deletion length profiles. These parameters are then learned from the data using a maximum-likelihood estimator, which is implemented in practice through an expectation-maximization algorithm. In this framework, the likelihood of a particular outcome of the SHM process is obtained by summing the probabilities of all possible scenarios of {\rev germline V templates}, point mutations, insertions, and deletions that would yield that sequence. In {\rev principle}, this means that all alignments to {\rev germline V segments} {\rev should be} taken into account, and not just the best-scoring one as in traditional methods. {\rev However, many {\rev unlikely} alignment scenarios can be pruned out of the list before computation (see Methods). In practice, this implies that only one plausible germline V gene is typically considered.} Once the parameters {\rev of the model} are learned, they can be analyzed in their own right, or the model may be used to generate synthetic {\rev datasets} to test hypotheses. Fig.~\ref{fig:synth_data_inference}A summarizes the general workflow of the method.

The generative model is defined so as to be simple enough to be tractable, while recapitulating the main features of the SHM process. Its parameters are directly interpretable in terms of basic SHM statistics.
Each sequence has a distinct point-mutation rate {\rev $\mu$ per base pair}, corresponding to its maturation age, drawn from a distribution $P(\mu)$. Motivated by the linear relation between indel and point-mutation rates (Fig.~\ref{fig:real_data_general_stats}C), we assume that the rates of insertion and deletion events are given by $\beta_{\rm ins}\times\mu$ and $\beta_{\rm del}\times \mu$, where $\beta_{\rm ins}$ and $\beta_{\rm del}$ are sequence-independent factors. Then insertions, deletions, and point mutations are drawn randomly with these three rates, uniformly along the sequence. The lengths of each insertion and deletion event are drawn from two distributions $P_{\rm ins}(\ell_{\rm ins})$ and $P_{\rm del}(\ell_{\rm del})$. The set of parameters to be inferred is collectively called $\theta=(P,\beta_{\rm ins},\beta_{\rm del}, P_{\rm ins}, P_{\rm del})$, and corresponds to the basic statistics of the SHM process. To calculate the sum of probabilities over all plausible scenarios efficiently, we use a forward algorithm that avoids computing the sum explicitly.
Mathematical details of the probabilistic model and of the likelihood maximization are reported in the Methods section.

\subsection*{Validation of the probabilistic approach on synthetic data}

The ability of our inference procedure to recover the true indel statistics can be tested by using large enough synthetic datasets. To this aim, we generated $\mathcal{N}=10$ independent repertoires of $100,000$ synthetic sequences each, restricted to the V segment, {\rev using realistic parameters similar to those obtained by the analysis of standard annotation (Fig.~\ref{fig:real_data_general_stats})}. Naive sequences were first generated by drawing V templates according to their frequencies in the data. Then, point and indel mutations were added according a generative SHM model belonging to the class described above. We assumed a mixture of $10\%$ naive cells, {\rev defined as cells that have not gone through the hypermutation process at all, $\mu=0$}, and $90\%$ experienced cells with $\mu$ distributed according to a shifted Gamma distribution mimicking a realistic distribution for point mutations (Gamma with mode $\mu=0.07$, standard deviation $=0.04$, and shift $-0.02$; {\rev see Methods for details}). The lengths of insertions and deletions followed geometric distributions with characteristic scale $\ell=15$ and a maximum value of 30 base pairs.

On each of these synthetic datasets, we compare two methods for estimating the parameters of the model.
The first method is the one used in the previous sections of the paper: synthetic sequences are aligned to germline V segments using a deterministic tool; then, from these alignments, the statistics $P(\mu)$, $\beta_{\rm ins}$, etc, are directly computed from their frequencies. The second method is the probabilistic inference approach outlined in the previous section.

In Fig.~\ref{fig:synth_data_inference}B we report results for a realistic {\rev rescaled} indel rate, $\beta_{\mathrm{del}}=\beta_{\mathrm{ins}}=0.025$, {\rev of the same order of magnitude of the ones found for nonproductive IgG sequences through standard annotation, see Fig.~\ref{fig:real_data_general_stats}A}. The fraction of naive cells ($\mu=0$), shown in inset, and the distribution of $\mu$ for experienced cells, shown in the main figure, are correctly inferred only by the probabilistic approach. An example of the error made by the deterministic approach involves naive sequences: sequences with no SHM are always labeled as ``naive'' by the best-scoring alignment, while the probabilistic method considers the possibility that {\rev they come} from an experienced cell ($\mu>0$) in which no SHM has had the chance to occur (which happens with probability $e^{-L\mu}$, where $L$ is the length of the V segment). This leads the deterministic method to systematically overestimate the true number of naive sequences.
The performance of the deterministic method degrades as the rescaled insertion rate $\beta_{\rm ins}$ is increased (and $\beta_{\rm del}$ with it, Fig.~\ref{fig:synth_data_inference}C). In that regime of frequent indels, best-scoring alignments merge nearby events, underestimating their relative frequency and compensating resulting errors by an increased mutation rate $\mu$.
By constrast, the probabilistic method recovers the ground truth even at very high indel densities, corresponding to $3$ deletion and $3$ insertion events per sequence on average.
The merging of nearby indels also causes the deterministic algorithm to find an excess of non-existing long insertion events above 30 base pairs, while the probabilistic one correctly assigns their frequency to zero (Fig.~\ref{fig:synth_data_inference}D). The analogous of panels~\ref{fig:synth_data_inference}C-D for deletions are reported in Fig.~\ref{fig:SI_synth_del}.

Overall, this analysis on synthetic data shows that while the deterministic analysis correctly captures the main statistics of SHM, it may fail on other quantities such as the fraction of truly naive cells, or the frequency of rare, long indels.

\subsection*{Inference within the probabilistic framework}

\begin{figure*}
\begin{center}
\includegraphics[width=\textwidth]{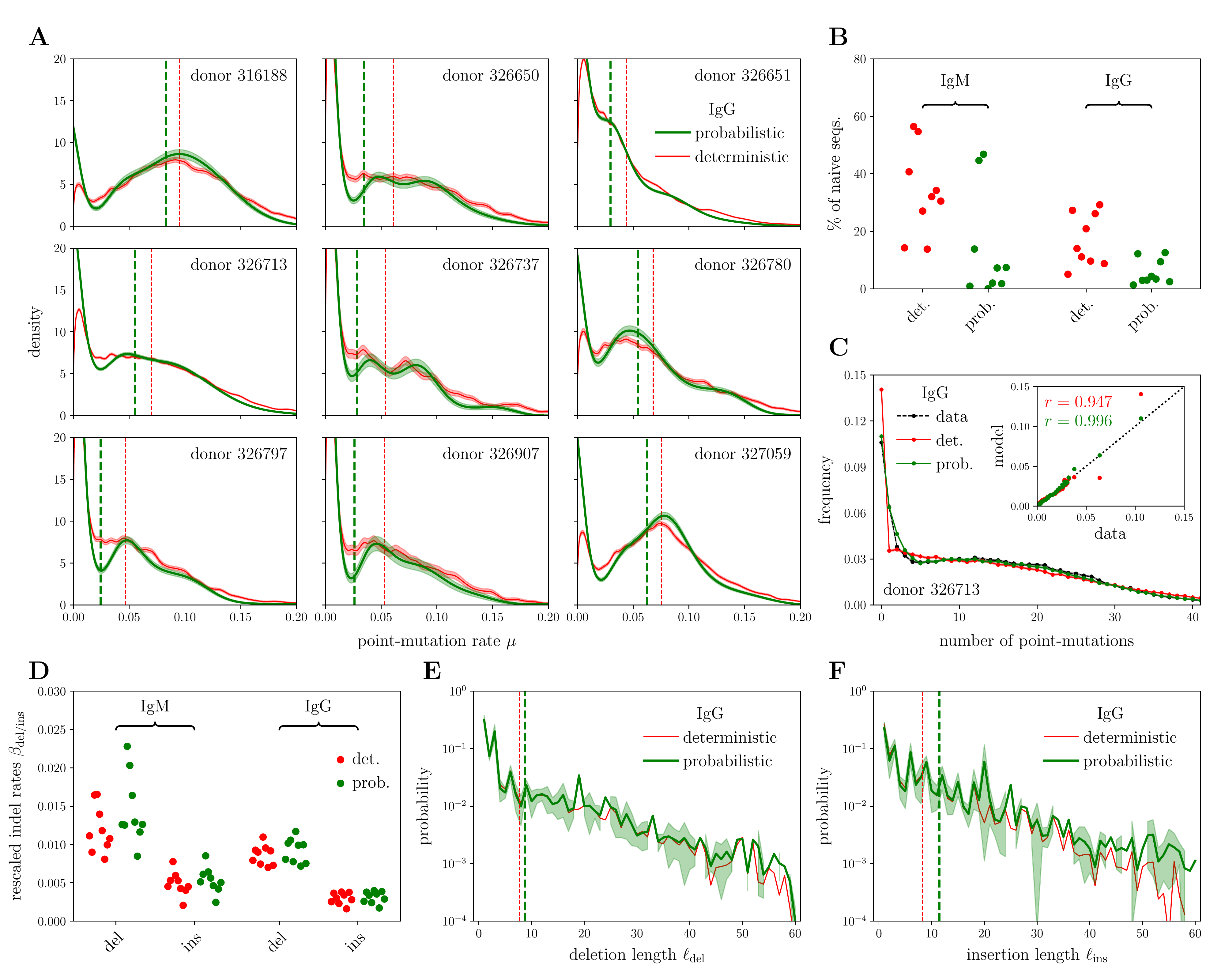}
\caption{\textbf{Results of probabilistic algorithm on IgG repertoire data}. Probabilistic SHM-indel models trained on the 9 Briney healthy donors~\cite{BrineyEtAl2019} (IgG sequences always shown, plus IgM only shown when relevant); for comparison, models based on deterministic assignments are also shown. \textbf{(A)} Point-mutation rate distributions; vertical lines locate average values for the distributions; deterministic distributions come from a smoothed version of SHM fraction histograms (more details in the main text); shaded areas show the confidence interval of the inference result (estimated from a Poisson noise assumption). \textbf{(B)} Corresponding estimates of the fraction of truly {\rev naive} sequences, with a comparison also with the IgM case. \textbf{(C)} Validation of the inference results on one of the largest donors (326713); full histogram for the number of SHM in the main plot, corresponding scatter plot in the inset (correlation coefficients: $r_{\mathrm{det.}}=0.947$ vs $r_{\mathrm{prob.}}=0.996$). \textbf{(D)} Estimates for the rescaled {\rev (i.e. divided by the point-mutation rate)} deletion and insertion rates per base pair; IgM is shown for comparison. \textbf{(E)} Inferred deletion length profile (solid line for the mean over the 9 donors, plus shaded area around for standard deviation), compared to the deterministic estimate (only mean shown). Vertical lines with the same color coding refer to the mean lengths. \textbf{(F)} Same as (E), but for insertions.}
\label{fig:real_data_inference}
\end{center}
\end{figure*}

We next applied our probabilistic inference method to the analysis of nonproductive IgH sequences from the 9 donors of \cite{BrineyEtAl2019} to learn the parameters of the SHM process free of selection effects.
We thus obtain individual-specific distributions of maturation ages (Fig.~\ref{fig:real_data_inference}A, green). For comparison, the naive deterministic estimate, based on the frequency of errors in best-scoring alignments, is also reported (red). We observe substantial variability across individuals, probably due to diverse histories of experience with antigens. The probabilistic method finds more sequences with a small, non-zero maturation age ({94.3\%$\pm$4.1\%}) than the deterministic one ({83.1\%$\pm$8.6\%}), see Fig.~\ref{fig:real_data_inference}B. Because such sequences often have no mutations, the deterministic method underestimates that number (as confirmed by our synthetic data results, see Fig.~\ref{fig:synth_data_inference}B), suggesting that the probabilistic approach gives a more accurate result.
While we don't know the ground truth for the true distribution of maturation ages, we can verify this interpretation on the distribution of point mutations.
In Fig.~\ref{fig:real_data_inference}C we compare predictions from the deterministic and probabilistic approaches for the IgG repertoire of one donor.
Consistent with our intuition, the deterministic distribution overrepresents unmutated sequences, while the probabilistic distribution is in excellent agreement with the data.
In addition, the model  reproduces the super-Poissonian distribution of the number of indels (Fig.~\ref{fig:SI_model_pred}).

In Fig.~\ref{fig:real_data_inference}D we report estimates for rescaled indel rates, i.e. ratios of indel to point-mutation rates.
They are consistent between the deterministic and probabilistic methods, as expected from the analysis on synthetic data, because of the relatively low indel rates in real data. They are also well conserved across individuals. While the raw mutation rates depend on the history of affinity maturation, ratios between them are not expected to depend on it. They should instead be determined by universal biophysical mechanisms of DNA damage and repair, which explains their relative conservation. Similarly, the indel length profiles (Figs.~\ref{fig:real_data_inference}E-F) are consistent between the deterministic and probabilistic methods, as well as across individuals, pointing again to the universality of the biophysical mechanisms at play. However, the average deletion and insertion lengths are slightly larger in the probabilistic estimate, due to a more reliable evaluation of low-probability events corresponding to the tails of the distributions.

\section{Discussion}
Most studies of SHM insertions and deletions, and even point mutations, consider productive rearrangements. Instead, we focused our analysis on nonproductive sequences to study the statistics of mutation inherent to the biochemistry of the SHM process, free of subsequent selection effects. However, we found a strong correlation between the productive and nonproductive positional profiles of SHM, implying that most of the heterogeneity is of biochemical origin, rather than stemming from selection. Further studies of the differences in SHM statistics between productive and nonproductive sequences could shed light on the structural and functional constraints that selection imposes on antibodies during affinity maturation.

Because nonproductive sequences and indel SHM are both rare, the combination of both is extremely rare, necessitating a very large dataset such as the one provided by {\em Briney et al.} \cite{BrineyEtAl2019}. This study included 9 subjects, which allowed us to assess inter-individual variability. We found that features of the SHM process associated to the biochemistry of the process, such as the positional profiles, the ratio of the number of indels to point mutations, and the distribution of insertion and deletion lengths, were mostly conserved across subjects. By contrast, the distribution of the mutational age of antibodies, which reflects their unique immune history, was specific to each individual. {\rev While we studied individual sequences, we know that those are organized in distinct lineages originating from naive germline sequences. This could lead to multiple-counting mutations occuring in several sequences of the same lineage. However, since this effect is random and decoupled from the mutation process itself in nonproductive sequences, we only expect it to increase errors due to sampling. The low individual variability reported for most summary statistics provides an upper bound on that error.}

We report a large positional heterogeneity, with a similar relative variability between point and indel hypermutations. Indels tend to concentrate around the CDR1 and CDR2 loops, as do point mutations \cite{SainiHershberg2015}, and consistently with previous reports on productive sequences \cite{BrineyEtAl2012}.
However, our results show that the positional profiles of point, insertion, and deletion SHM are still distinct (albeit correlated), suggesting a complex picture of how the 3 processes result from AID selectivity and positional preference.

The precise mechanism causing SHM indels and point mutations is still elusive. Our results confirm the main prediction of the polymerase slippage model~\cite{StreisingerEtAl1966, GoldingEtAl1987} for indels, namely that inserted segments are homologous to their immediately flanking regions. This complements earlier observations from \cite{BrineyEtAl2012}, but without the interference of selection, and on a much bigger dataset allowing for more detailed statistics.
By contrast, we found that deleted segments are not significantly homologous to their neighborhood, suggesting the slippage is not driven by sequence homology, but reaches out to a random position along the DNA. {\rev An alternative hypothesis is that deletions are driven by double-strand breaks caused by AID, consistent with our and previous observations that they tend to co-localize with point mutations \cite{YeapEtAl2015,HwangEtAl2017}}. In the slippage model, the indel length corresponds to the distance by which the polymerase slips. Both have a characteristic exponential tail in their distribution, compatible with the processive motion of the slipping polymerase with random stopping.

Our probabilistic approach is not only useful for annotating sequences and {\rev inferring} rates, but it can also be used as a generative model to produce synthetic datasets for computational controls. However, the model makes a few simplifying assumptions for the sake of computational tractability. First, its positional profiles are constant, in contradiction with the data. Second, inserted segments are assumed to be uniformly random, ignoring homology to flanking regions. While these approximations are unlikely to affect our results, future refinements of the generative model would be welcome to create more realistic datasets.

We evaluated the conditions under which the probabilistic approach we introduced added value to the classical deterministic annotation of mutated antibodies. For most of the bulk repertoire, where the rate of hypermutation is relatively low, both approaches yielded similar results, with a notable exception: the fraction of {\rev ``naive''} sequences that have not undergone any affinity maturation.
{\rev This definition of naive is distinct from usual phenotypic characterization based on isotype classification, flow cytometry \cite{GlassEtAl2020} or single-cell RNA sequencing data \cite{HornsEtAl2020}, although they are correlated.}
This proportion is always overestimated by the deterministic method. Since the SHM process is stochastic, some sequences having undergone affinity maturation may not have accumulated any mutation at all. These sequences are always considered naive by the deterministic approach, while the probabilistic method considers both possibilities, correcting that bias.

Beyond the bulk of the repertoire, the probabilistic approach is most useful for analyzing heavily mutated sequences, where it is difficult to distinguish point mutations from nearby insertion and deletion events. While a minority, such sequences are very important. They are found in the largest clonal families having undergone the longest affinity maturation, and so are the most experienced and selected ones. HIV1-related bnAbs, which often contain many indels, are examples of such sequences \cite{SokBurton2018}. The relatively low frequency of indels that we report could explain why effective bnAbs are rare, and only emerge naturally in a small percentage of infected individuals. In that regard, evidence that indels frequently appear in response to some vaccination protocols~\cite{ReasonZhou2006} suggests promising strategies for HIV vaccination.

\section{Methods}
\subsection*{Data and preprocessing}

We exploited publicly available IgM and IgG repertoire data from a recent ultra-deep repertoire study of immunoglobulin heavy chains (IgH), including approximately $3$ billions sequences in total, coming from 9 different healthy donors~\cite{BrineyEtAl2019}. The V region is well covered by the mRNA sequencing protocol, starting from nucleotide position $\sim 66$ along the templates for almost all the sequences; CDR3 and J gene are also covered.

We first preprocessed raw sequences as in~\cite{SpisakEtAl2020}, using pRESTO {\rev (release 0.5.13)} from the Immcantation pipeline~\cite{VanderHeidenEtAl2014}. Then, we performed a standard annotation through the IgBLAST standalone tool~\cite{YeEtAl2013}, {\rev (release 1.13.0, with default values for penalties)}, referring to IMGT for germline templates~\cite{GiudicelliEtAl2006} ({\rev only functional (F) genes are considered, totaling 265 IGHV different templates, alleles included}). We also applied quality filters, keeping only sequences with properly annotated V gene, J gene and CDR3, and covering at least $200$ base pairs of the V segment. {\rev We checked using synthetic data that this quality filter does not substantially alter the underlying indel distribution below length 60, although sequences with very long indels are slightly less likely to pass it (Fig.~\ref{fig:SI_quality_filter}).}

Unique molecular identifiers (UMI) were used in the original study to correct for PCR amplification biases. We grouped reads with the same UMI to build consensus sequences for downstream analysis in which read counts were ignored, {\rev allowing for a maximum error rate of 10\% within each group. Sequences corresponding to low numbers of reads per UMI (e.g. 1 or 2) are expected to be more affected by sequencing errors, because of the inability to build a consensus sequence from multiple reads. We verified that our results are not affected by this potential source of error, by repeating the standard annotation analysis on UMIs represented by at least 3 reads (Fig.~\ref{fig:SI_UMI_check}). Thus, to make the most of the available data, we did not discard sequences with low number of reads per UMI.}

It is very hard to call mutations in the CDR3 because of its variability, and the J region is very short. Therefore we restricted ourselves to the V region, starting from the beginning of the read, up to the conserved cysteine before the beginning of the CDR3. We call that sequence $s$.

\subsection*{Modeling approach}

{\rev We start from a repertoire $\{s\}$ of size $N$. Let $\{t\}$ be the set of (truncated) V templates used for the alignment (the same 265 templates exploited for IgBLAST annotation). For each $s$ and $t$, a deterministic alignment algorithm, based on the standard Needleman-Wunsch (NW) algorithm for pairwise global alignments with affine score for gaps~\cite{Book_DurbinEtAl1998}, provides \textit{only} the alignment with the highest score $\mathcal{S}(s,t)$.

  The value of the parameters used in the NW algorithm were chosen according to the actual frequency of related events in human IgG heavy chains, as estimated through IgBLAST annotation (Fig.~\ref{fig:real_data_general_stats}). Assuming an average point-mutation rate of $\sim 5\%$, an indel rate of $0.05\%$, and an exponentially decaying length distribution of indels with characteristic length $\ell=10$, we have the following log-odd penalties (using natural logarithm): $\simeq -0.05$ for matching and $\simeq -3$ for mismatching aligned base pairs; $\simeq -7.7$ for opening a deletion gap; $\simeq -0.1$ for extending it; $\simeq -9.1$ for opening an insertion gap; and $-1.5$ for extending it (including the $\log(4)\approx 1.4$ penalty for chosing the base pair)~\cite{Book_DurbinEtAl1998}. Note that since the alignment is global, only relative values of the penalty matter.}

The optimal alignment is built through a dynamic-programming strategy, relying recursively on the optimal alignments of shorter blocks. The key modification we introduce to the standard NW algorithm is to sum over all the possibilities during the recursive scheme, rather than choosing the optimal one at each step. The result is then no longer an optimal global alignment score, but rather the sum of the likelihoods of all the possible global alignments between~$s$ and~$t$, summed over all templates~$t$. Each possible alignment contributes to the final sum proportionally to its likelihood, yielding a {\em probabilistic} likelihood of the sequence, by contrast with the {\em deterministic} score of the standard NW algorithm.

The resulting likelihood of the sequence $s$, $\mathcal{L}(s;\theta)$, depends on the model parameters $\theta$.
The total likelihood of the whole repertoire is then given by:
\begin{equation}
	\mathcal{L}(\theta) = \prod_s \mathcal{L}(s;\theta),
	\label{eq:total_L_def}
\end{equation}
where we assumed independent sequences, {\rev ignoring the lineage structure of the repertoire (see Discussion)}. Maximizing the likelihood with respect to $\theta$ gives an estimate~$\hat{\theta}$ for the alignment parameters that best describe the data.

\subsection*{Recursive algorithm for the alignment}

We now give the details of the computation of $\mathcal{L}(s;\theta)$. Consider a sequence $s$ and a template $t$. Indexing starts from zero, including an initial ``placeholder'' position for potential gaps, running over actual base pairs from $s$ and $t$ (gaps are not taken into account directly in this indexing). So $s_i$ is the $i$-th symbol along $s$, while $s_{0:i}$ is the portion of $s$ running from the first actual symbol $s_1$ to $i$-th symbol $s_i$ included, preceded by the ``empty'' placeholder position 0.

As in NW, the alignment procedure always starts from the beginning of the two sequences $s$ and $t$, and goes through all the possibilities by extending the previous alignment for shorter pieces of $s$ and $t$. In our case, however, we sum over these possibilities, rather than keep the best-scoring one. The indices on $s$ and $t$ are not constrained to increase at the same time; an asymmetric increase corresponds to a deletion or an insertion.

The alignment score for the portion $s$ up to position $i$ included ($s_{0:i}$) and the portion of $t$ up to position $j$ included ($t_{0:j}$) is computed by relying on the previous alignments of shorter portions, according to the following possibilities:
\begin{itemize}
	\item $s_{i-1}$ and $t_{j-1}$ were previously aligned to each other, so the alignment advances by one position on both sequences, with a \textit{match} ($s_i = t_j$) or a \textit{mismatch} ($s_i \neq t_j$);
	\item $t_j$ was previously aligned with $s_k$, $k<i$, so that a gap on $s$ of length $i-k$ has to be taken into account (i.\,e., a \textit{deletion} of templated bases);
	\item $s_i$ was previously aligned with $t_{k'}$, $k'<j$, so that a gap on $t$ of length $j-k'$ has to be taken into account (i.\,e., an \textit{insertion} of non-templated bases).
\end{itemize}
In the standard NW algorithm, each of these possibilities is linked to a score, and only the largest among them is kept. Here, instead, we work directly with likelihoods, and then we sum over them according to the following recursion:
\begin{equation}
\begin{aligned}
	\mathcal{S}(s_{0:i},t_{0:j}) = &\,\mathcal{S}(s_{0:i-1},t_{0:j-1}) \, M(s_i,t_j)\\
	&+\sum_{k<i}\mathcal{S}(s_{0:k},t_{0:j}) \, \Gamma_{\text{del}}(i-k)\\
	&+\sum_{k'<j}\mathcal{S}(s_{0:i},t_{0:k'}) \, \Gamma_{\text{ins}}(j-k')
\end{aligned}
	\label{eq:fullS}
\end{equation}
where $M(s_i,t_j)$ is the match/mismatch probability between nucleotide $s_i$ and $t_j$, $\Gamma_{\text{del}}(\ell)$ is the probability of a deletion of length $\ell$, and $\Gamma_{\text{ins}}(\ell)$ is the probability of an insertion of length $\ell$. The initial conditions of the recursion are:
\begin{equation}
	\begin{aligned}
		&\mathcal{S}(s_0,t_0) &&= &&1\\
		&\mathcal{S}(s_{0:i},t_0) &&= &&\sum_{k<i}\mathcal{S}(s_{0:k},t_0) \, \Gamma_{\text{del}}(i-k)\\
		&\mathcal{S}(s_0,t_{0:j}) &&= &&\sum_{k'<j}\mathcal{S}(s_0,t_{0:k'}) \, \Gamma_{\text{ins}}(j-k').
	\end{aligned}
\end{equation}
The final score $\mathcal{S}(s,t)\equiv\mathcal{S}(s_{0:L_s},t_{0:L_t})$ gives the likelihood of $s$ aligning to $t$, obtained as the sum over all possible alignments.

Likelihoods $M$, $P_{\rm del}$ and $P_{\rm ins}$ can be further described as:
\begin{equation}
	\begin{aligned}
		&\Gamma_{\text{del}}(\ell) \equiv \mu_{\text{del}}\,\times\,P_{\text{del}}(\ell)\\
		&\Gamma_{\text{ins}}(\ell) \equiv \mu_{\text{ins}}\,\times\,P_{\text{ins}}(\ell)\,\times\,(1/4)^{\ell},
	\end{aligned}
\end{equation}
assuming uniformly random insertions, where $P_{\text{del}/\text{ins}}(\ell)$ is the distribution of deletion and insertion lengths, and:
\begin{equation}
	\begin{aligned}
		M(s_i,t_j) \equiv &\,(1-\mu_{\text{del}}-\mu_{\text{ins}}) \\
		&\times \, \left[(1-\mu)\,\mathbb{I}(s_i=t_j)+\mu\,\mathbb{I}(s_i \neq t_j)\right]
	\end{aligned}
\end{equation}
with $\mathbb{I}(\cdot)$ being the indicator function (equal to one if argument is true, otherwise equal to zero). The rates $\mu_{\rm del}$, $\mu_{\rm ins}$, and $\mu$ are interpreted as the deletion, insertion, and point mutation rates {\rev per base pair}.

Based on evidence from data (Fig.~\ref{fig:real_data_general_stats}C), we assume that these rates scale together with the mutational age, since both types of mutations accumulate with time during affinity maturation cycles. We thus write:
\begin{equation}
\mu\equiv \mu_s,\quad	\mu_{\text{del}} \equiv \mu_s \times \beta_{\text{del}} \quad , \quad \mu_{\text{ins}} \equiv \mu_s \times \beta_{\text{ins}},
\end{equation}
where $\beta_{\text{del}}$ and $\beta_\text{ins}$ are constant (repertoire-specific) ratios, and $\mu_s$ is sequence specific, as suggested by the wide range of mutabilities in the data. The ratios $\beta_{\text{del}/\text{ins}}$ will also be referred to as \textit{rescaled} deletion and insertion rates.
The mutation rate $\mu_s$ is treated as a hidden variable, whose distribution $P(\mu)$ is a model parameter.

The likelihood of each sequence is obtained by summing over all mutation rates and templates:
\begin{align}\label{eq:Ls}
  \mathcal{L}(s;\theta)&=\int d\mu_s\,P(\mu_s)\mathcal{L}(s|\mu_s;\phi),\\
  \textrm{with }\mathcal{L}(s|\mu_s;\phi)&\equiv \sum_t p_t\,\mathcal{S}(s,t|\mu_s;\phi)\label{eq:sumt},
\end{align}
with $p_t$ uniform, and where the parameters $\theta$ have been separated into two components: $\theta=(P(\mu),\phi)$, with $\phi\equiv (\beta_{\rm del},\beta_{\rm ins}$,$P_{\rm del}(\ell)$,$P_{\rm ins}(\ell))$. {\rev In Eq.~(\ref{eq:sumt}) we made explicit the dependence of alignment score on the mutational age $\mu_s$ and indel parameters $\phi$.}

The repertoire is composed of a mixture of cells that have undergone some mutational process, presumably in germinal centers, and cells that have stayed naive. To properly model this, we further assume that the prior distribution of mutation rates has two parts:
\begin{equation}
	P(\mu) = f \, \delta(\mu) + (1-f) \, \tilde{P}(\mu),
	\label{eq:generalized_prior}
\end{equation}
where $f$ is the fraction of naive cells, and $\tilde{P}(\mu)$ is a smooth probability density.

\subsection*{Likelihood maximization}

To maximize the likelihood, Eqs.~(\ref{eq:total_L_def}) and (\ref{eq:Ls}), we employ a combination of Expectation-Maximization (EM) \cite{DempsterEtAl1977, Book_McLachlanKrishnan2008} for updating $P(\mu)$, and gradient ascent for updating $\phi$.

We denote by $\theta^t=(P^t(\mu),\phi^t)$ the value of the parameters at iteration $t$. Following EM, we define the pseudo log-likelihood to be maximized with respect to $\theta$ at each iteration:
\begin{equation}
\theta^{t+1} \leftarrow \mathrm{arg}\max_{\theta} Q(\theta|\theta^t),
\end{equation}
with
\begin{equation}
		Q(\theta|\theta^t) = \sum_s \left<\log[{\mathcal{L}(s|\mu_s;\phi)\,P(\mu_s)}]\right>_{\mu_s|s;\theta^t},
\label{eq:Q}
\end{equation}
where the mean $\left<\cdot\right>_{\mu_s|s;\theta^t}$ is with respect to the posterior distribution:
\begin{equation}
	P(\mu_s|s;\theta^t) \equiv \frac{\mathcal{L}(s|\mu_s;\phi^t)\,P^t(\mu_s)}{\int d\mu_s \, \mathcal{L}(s|\mu_s;\phi^t)\,P^t(\mu_s)}.
	\label{eq:posterior_def}
\end{equation}
Because of the separation between naive $\mu_s=0$ and experienced cells, this posterior may also be decomposed as:
\begin{equation}
	P(\mu_s|s;\theta^t) = f_s^t \, \delta(\mu_s) + (1-f_s^t) \, \tilde{P}(\mu_s|s;\theta^t),
	\label{eq:generalized_posterior}
\end{equation}
with
\begin{equation}
f_s^t=\frac{f^{t}\mathcal{L}(s|0;\phi^{t})}
{f^{t}\mathcal{L}(s|0;\phi^{t})+
(1-f^{t})\int d\mu_s \, \mathcal{L}(s|\mu_s;\phi^{t}) \, \tilde{P}^{t}(\mu_s)},
\end{equation}
and
\begin{equation}
	\tilde{P}(\mu_s|s;\theta^{t}) = \frac{\mathcal{L}(s|\mu_s;\theta^{t}) \, \tilde{P}^{t}(\mu_s)}{\int d\mu_s \, \mathcal{L}(s|\mu_s;\theta^{t}) \, \tilde{P}^{t}(\mu_s)}.
\end{equation}

The pseudo-likelihood $Q$ can be decomposed into two terms corresponding to the two factors inside the logarithm of Eq.~(\ref{eq:Q}), $\log \mathcal{L}(s|\mu_s;\phi)$ and $\log P(\mu_s)$. The first one only depends on $\theta$ through $\phi$, and the second one only through $P(\mu)$, allowing for their independent maximization.

\subparagraph*{Maximization with respect to $P(\mu)$.} To maximize $Q$ under the normalization constraint, we define the functional with Lagrange multipliers:
\begin{equation}
	\widehat{Q}(\theta|\theta^t) \equiv Q(\theta|\theta^t) + \lambda\left(1-\int d\mu \, P(\mu)\right).
\end{equation}
The extremal condition with respect to $P(\mu')$ gives:
\begin{equation}
	0 = \frac{\delta \, \widehat{Q}}{\delta \, P(\mu')} = \sum_s \int d\mu_s \, P(\mu_s|s;\theta^t) \, \frac{\delta(\mu_s-\mu')}{P(\mu_s)} - \lambda
\end{equation}
When plugging functional forms (\ref{eq:generalized_posterior}) and (\ref{eq:generalized_prior}) in it, together with the extremal condition with respect to $\lambda$, we get:
\begin{align}
	\tilde{P}^{t+1}(\mu) & \leftarrow \frac{\sum_s (1-f^t_s) \, \tilde{P}(\mu|s;\theta^t)}{\sum_s (1-f^t_s)}\\
	f^{t+1}  & \leftarrow  f = \frac{1}{N}\sum_s f^t_s.
\end{align}

Intuitively, the new estimate of the smooth part $\tilde{P}(\mu)$ of the repertoire-wide distribution is the weighted average of the sequence-specific posterior distributions over the repertoire. Similarly, the new estimate for the fraction $f$ of naive cells is given by the average over the repertoire of the probability of being naive.

\subparagraph*{Maximization with respect $\phi$.} To maximize the first term of $Q$
with respect to $\phi$,
we use Monte Carlo sampling to represent the integral over the posterior $P(\mu_s|s;\theta^t)$:
\begin{equation}
		\sum_s \left<\log{\mathcal{L}(s|\mu_s;\phi)}\right>_{\mu_s|s;\theta^t} \simeq \sum_s \sum_{n=1}^{N_{\mathrm{MC}}} \log{\mathcal{L}(s|\mu_{s,n};\phi)},
\end{equation}
with $\mu_{s,n} \sim P(\mu_s|s;\theta^t)$.

Then, the extremal condition with respect to $\phi_i$ reads:
\begin{equation}
	\nabla^t_i\equiv \frac{\partial}{\partial \phi_i}\sum_s \sum_n \log{\mathcal{L}(s|\mu_{s,n};\phi)}=0.
\end{equation}
Since we could not find an exact update rule for $\phi_i$, we relied on gradient ascent to maximize the likelihood. Each gradient component $\nabla^t_i$ can be estimated numerically by infinitesimally shifting $\phi_i$:
\begin{equation}
		 \nabla^t_i\simeq \sum_s\sum_n \frac{\log{\mathcal{L}_s(s|\mu_{s,n};\tilde{\phi}^t_i)}-\log{\mathcal{L}_s(s|\mu_{s,n};\phi^t)}}{\varepsilon},
\end{equation}
where $\tilde{\phi}^t_i$ is equal to $\phi^t$ for all the components but $i$, which has been increased by $\varepsilon$.

We updated $\phi$ according to a momentum gradient-ascent update rule:
\begin{equation}
	\phi^{t+1} \,\, \leftarrow \,\, \mathcal{P}\left(\phi^t + \alpha^t\nabla^t + \omega^t(\phi^t-\phi^{t-1})\right),
	\label{eq:phi_update}
\end{equation}
with learning rate $\alpha^t$ and inertial term $\omega^t$, whose dependence on the iteration step $t$ was chosen to optimize convergence and to avoid long-time oscillations~\cite{ParikhBoyd2014}:
\begin{equation}
	\alpha^t \propto e^{-2t/T} \quad , \quad \omega^t \propto \frac{t}{t+3},
\end{equation}
with $T$ being the maximum number of time iterations allowed. {\rev In Eq.~(\ref{eq:phi_update}), $\mathcal{P}$ denotes the projection onto the simplex defined by the constraints $\beta_{\text{del}}\geq 0$, $\beta_{\text{ins}}\geq 0$, $P_{\rm del}(\ell)\geq 0$, $P_{\rm ins}(\ell)\geq 0$, and $\sum_\ell P_{\rm del}(\ell)=\sum_\ell P_{\rm ins}(\ell)=1$. Projection was done using the procedure described in Ref.~\cite{DuchiEtAl2008}}.

\subsection*{Speeding up the computation}
The algorithm described so far is computationally very costly.
Just the basic step of computing the alignment likelihood $\mathcal{L}(s|\mu_s;\phi) $ for a single sequence at fixed $\mu_s$ is time-consuming: if we allow for a maximum size $\ell=\Theta$ for single-event deletions and insertions, then the requested number of operations roughly scales as $L_s \times \left<L_t\right> \times N_t \times (2\Theta+1)$, where $N_t$ is the number of V templates considered and $\left<L_t\right>$ is their average length. This has to be repeated for each sequence, for each mutation rate, for each template, and for each direction of the gradient, and all of this at each iteration.
Below we describe approximations that considerably speed up the code.

\subparagraph*{Pruning the alignment matrix.} Indels are quite rare in real Ig sequences, with most sequences having none or at most one, located around well defined regions. Allowing for possible indels (of size $\Theta$ at most) everywhere in every sequence contributes very little to the final cumulated alignment likelihood, wasting computational time.
Dropping implausible terms would dramatically reduce the computational cost, pushing the factor $L_s \times \left<L_t\right> \times (2\Theta+1)$ almost down to $L_s$.

Plausible locations for gaps can be identified by computing, for each pair of positions $(i,j)$ along $s$ and $t$, the likelihood that the alignment goes through that pair. This is given by:
\begin{equation}
  \mathcal{S}_{ij} \equiv \mathcal{S}(s_{1:i}, t_{1:j}) \times \mathcal{S}(s_{i+1:L_s}, t_{j+1:L_t}),
  \end{equation}
where the first term is computed as described above, and the second term similarly, but using a backward iteration on the reverted sequences. Normalizing by the total alignment likelihood, we obtain a relative probability that the correct alignment passes through $(i,j)$, $\hat{\mathcal{S}}_{ij} = \mathcal{S}_{ij}/\mathcal{S}(s,t)$.

The entries of this matrix can now be used to \textit{prune} the terms of the recursive Eq.~(\ref{eq:fullS}). When $\hat{\mathcal{S}}_{i,j}$ is smaller than a fixed threshold $\vartheta=10^{-5}$, we set $\mathcal{S}(s_{0:i},t_{0:j})$ to zero and avoid computing it and including it in further sums.

To implement this strategy, we need to first run the forward and backward iterations without any pruning, for an initial guess of the parameters $(\mu_s,\phi)$. The $\hat{\mathcal{S}}_{i,j}$ matrix computed in this way is then used to define the pairs of positions to be kept in future computations.
After that, all subsequent evaluations of $\mathcal{S}(s,t)$ in the optimization algorithm {\rev are} done with only those pairs of positions.

\subparagraph*{Reducing the number of templates.} Another important performance improvement comes from pruning the list of plausible templates for each sequence $s$. In principle, the sum in Eq.~(\ref{eq:sumt}) runs over all the templates~$t$, but in practice only one of them actually contributes in a non negligible way. Thus, we only keep the V template $t^*_s$ closest to $s$, as given by the alignment score of the standard NW algorithm:
\begin{equation}
	\mathcal{L}(s|\mu_s;\phi) = \sum_t p_t \, \mathcal{S}(s,t|\mu_s;\phi) \simeq \mathcal{S}(s,t^*_s|\mu_s;\phi).
\end{equation}
The computational cost of $\mathcal{L}(s;\theta)$ is then reduced further by a factor $N_t$.

\subparagraph*{Averaging over the hidden age factor.} Another demanding operation is the average over the hidden variable $\mu_s$ by Monte-Carlo sampling. In principle, the number of Monte-Carlo samples $N_{\rm MC}$ should be very large. In practice, we used a single $N_{\rm MC}=1$ sample at each iteration. This simplification works because parameters evolve slowly at each iteration, especially in the last convergence steps, which allows for time averaging as a substitute for large MC sampling.

\subparagraph*{Posterior approximation.} To calculate and sample from the posterior $P(\mu_s|s;\theta)$, we evaluated it on a finite set of values of $\mu$ (called nodes), and interpolated the rest of the function using a smooth piecewise cubic interpolation~\cite{KlugeSpline2015}. The nodes are set for each sequences to be placed around the naive estimate of the mutation rate $\mu_s$ based on the NW alignment, in addition to 2 key nodes at $\mu=0$ and $1$. We used from 15 to 25 nodes depending on how mutated the sequence was.

As both alignment parameters $\phi$ and repertoire-wide distribution $P(\mu)$ get updated during inference, the positions of the nodes are updated for each sequence as well, by placing them around the previously estimated maximum of the posterior with a spread given by its inverse curvature.

The prior {\rev distribution} $P(\mu)$ was numerically stored by binning uniformly the $[0,1]$ interval with a bin size of $0.0005$.

{\rev
\subsection*{Generation of synthetic data}
In order to mimic real distributions of the repertoire-wide point-mutation rate $P(\mu)$, we used a \textit{shifted} Gamma distribution:
\begin{equation}
	P(\mu) = \frac{1}{z(\alpha,\beta,\mu_0)}(\mu+\mu_0)^{\alpha-1}e^{-\beta \mu}
\end{equation}
with $\alpha,\beta$ determined by the mode and standard deviation of the unshifted Gamma distribution: $(\alpha-1)/\beta=0.07$ and $\sqrt{\alpha}/\beta=0.04$; and where $z(\alpha,\beta,\mu_0)$ is a normalization constant. In practice, we drew values from the Gamma distribution with parameters $\alpha,\beta$, and then subtracted $0.02$ from it, discarding negative values.
The resulting distribution has non-zero probability for infinitesimally small values and a mode of $\mu=0.05$. We used this distribution for generating synthetic data when validating the software and reported as ground truth in Fig.~\ref{fig:synth_data_inference}.
}

\section*{Data availability}
All the data analyzed in this paper has been previously published and can be accessed from original publications. The software here presented is freely available at \url{https://github.com/statbiophys/indie}.

\section*{Acknowledgements}
The study was supported by the European Research Council COG 724208. 

\bibliographystyle{pnas}

\onecolumngrid

\newpage

\section*{Supplementary information}

\renewcommand{\thefigure}{S\arabic{figure}}
\setcounter{figure}{0}
\setcounter{section}{0}

\begin{figure*}[h]
\begin{center}
\includegraphics[width=\textwidth]{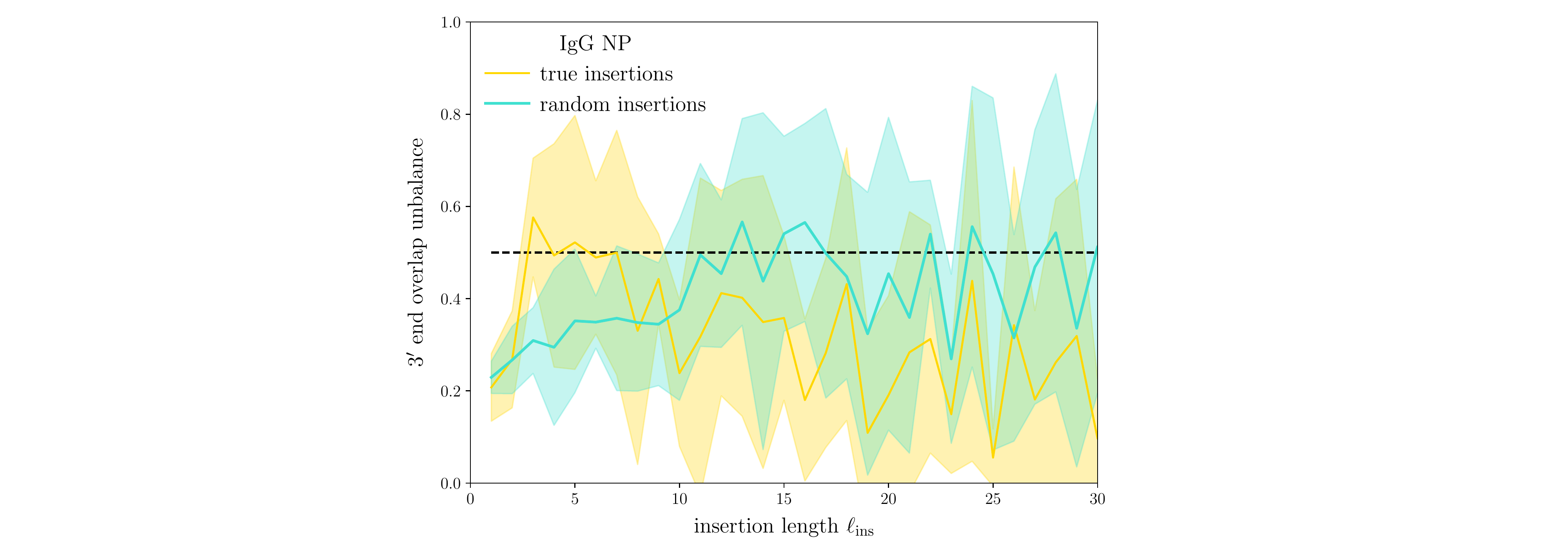}
\caption{Fraction of times the overlap between inserted base pairs and same-length flanking region on the $3'$ {end} is larger than the overlap with the $5'$ {end} along the sequence; averages and one standard deviation over the 9 Briney donors given by solid line and shaded area around, respectively. For comparison, overlap between random insertions and flanking regions is also reported (details in the main text). A weak preference for the $5'$ {end} is supported by a KS-test p-value of $0.04$, when comparing true insertions with randomized ones.}
\label{fig:SI_overlap_ins}
\end{center}
\end{figure*}

\begin{figure*}[h]
\begin{center}
\includegraphics[width=\textwidth]{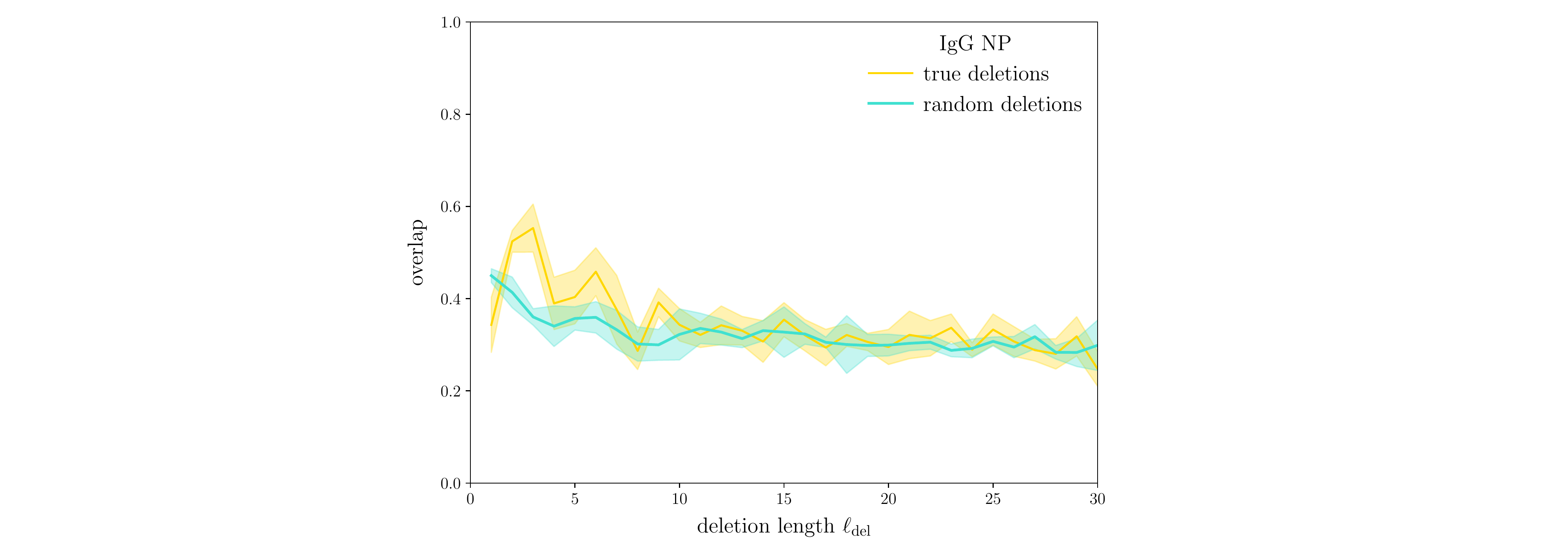}
\caption{Overlap between deleted base pairs and same-length flanking regions on either $3'$ or $5'$ {end} along the sequence (larger is kept); averages and one standard deviation over the 9 Briney donors given by solid line and shaded area around, respectively. For comparison, overlap between random deletions and flanking regions is also reported (details in the main text).}
\label{fig:SI_overlap_del}
\end{center}
\end{figure*}

\begin{figure*}[h]
\begin{center}
\includegraphics[width=\textwidth]{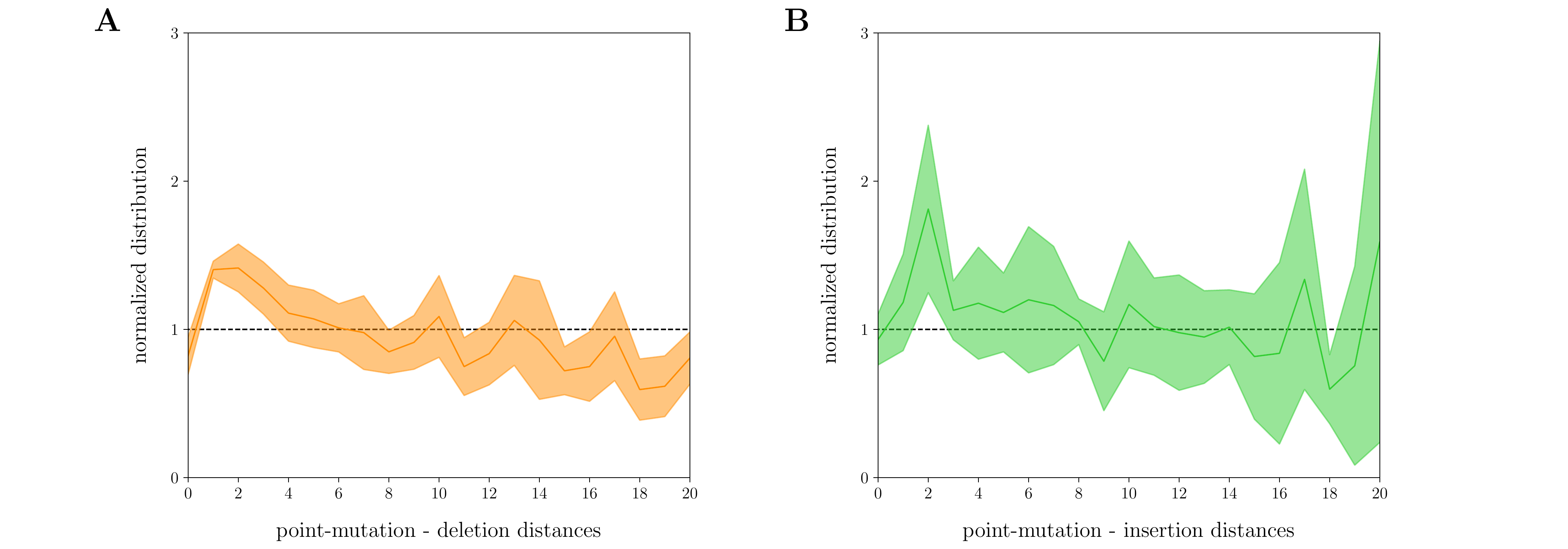}
\caption{\rev Distribution of the distance (in base pairs) separating
  \textbf{(A)} deletions or \textbf{(B)} insertions from the closest
  point mutation in nonproductive IgG sequences, normalized by the
  null expectation obtained by reshuffling indel and point mutations
  between sequences with the same V gene (to control for the positional biases of Fig.~\ref{fig:real_data_V_profiles}). Average and error bars are over the 9 donors.}
\label{fig:SI_corr_SHM_del}
\end{center}
\end{figure*}

\begin{figure*}[h]
\begin{center}
\includegraphics[width=\textwidth]{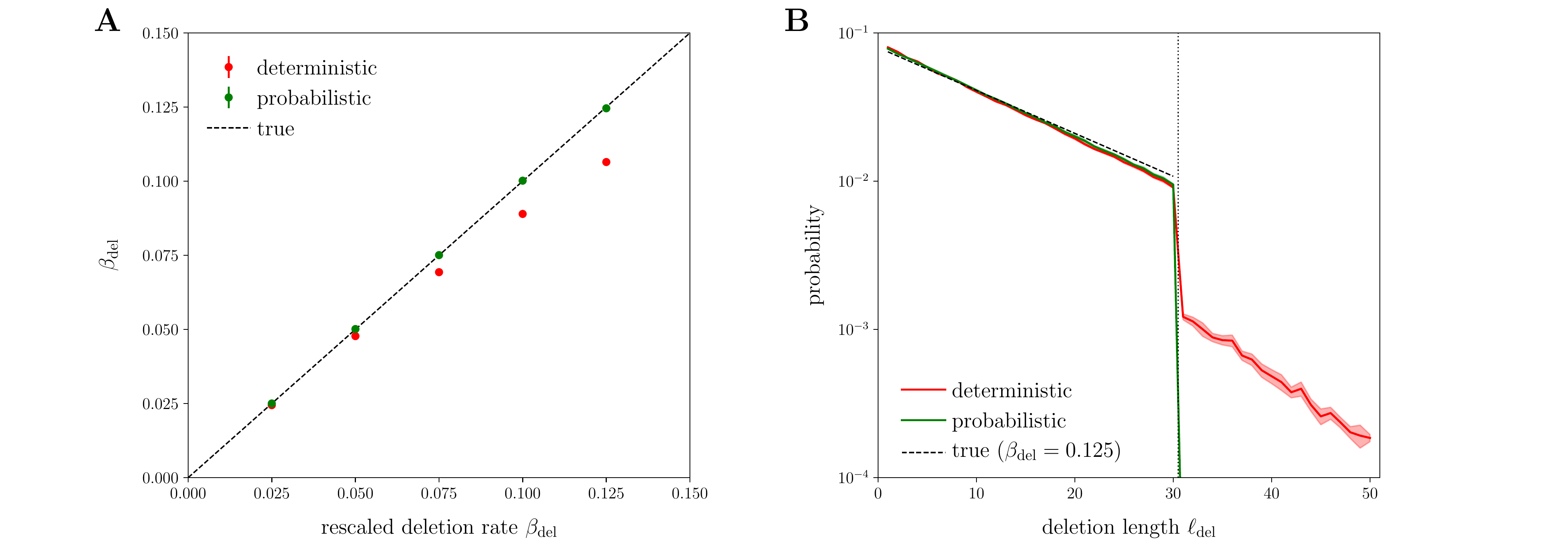}
\caption{\textbf{(A)} Probabilistic vs deterministic estimates of the rescaled deletion rate $\beta_{\rm del}$ at increasing values of indels density. Averages values plus one standard deviation error bars are obtained over the $\mathcal{N}$ independent synthetic repertoires. \textbf{(B)} Deletion length profiles for the largest value of $\beta_{\rm del}$ considered in panel (A); mean and standard deviations over the $\mathcal{N}$ independent realizations.}
\label{fig:SI_synth_del}
\end{center}
\end{figure*}

\begin{figure*}[h]
\begin{center}
\includegraphics[width=\textwidth]{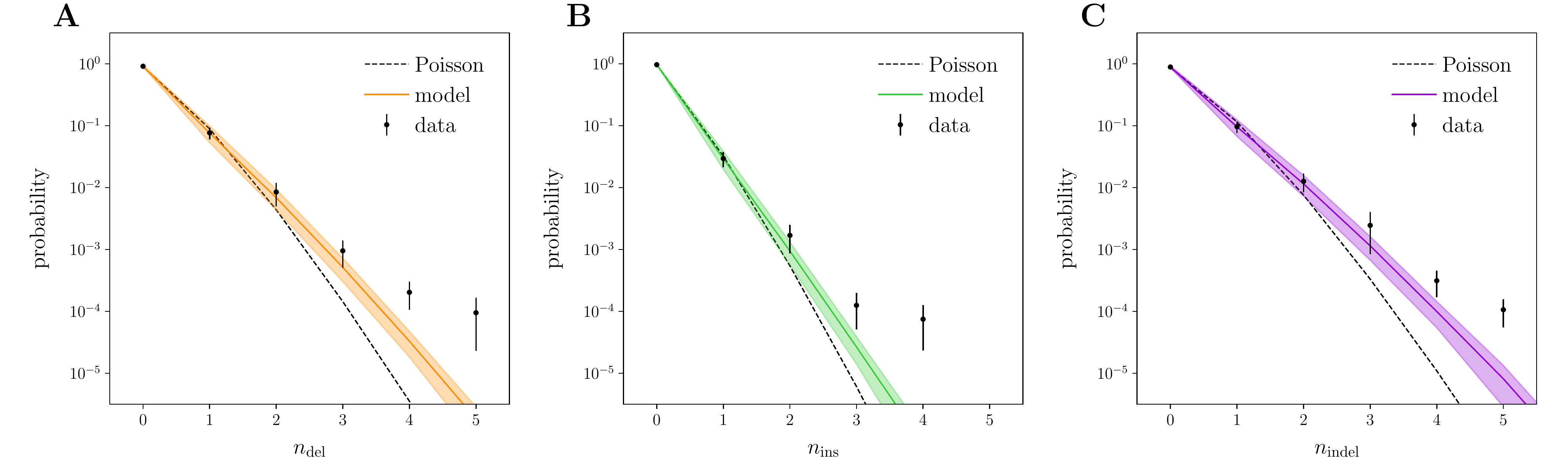}
\caption{Model prediction for frequency of \textbf{(A)} deletions, \textbf{(B)} insertions, and \textbf{(C)} both insertions and deletions, in IgG nonproductive sequences, compared to the data. Mean and variance are over the 9 donors. Poisson distributions with the same means as the data are shown for comparison.}
\label{fig:SI_model_pred}
\end{center}
\end{figure*}

\begin{figure*}[h]
\begin{center}
\includegraphics[width=\textwidth]{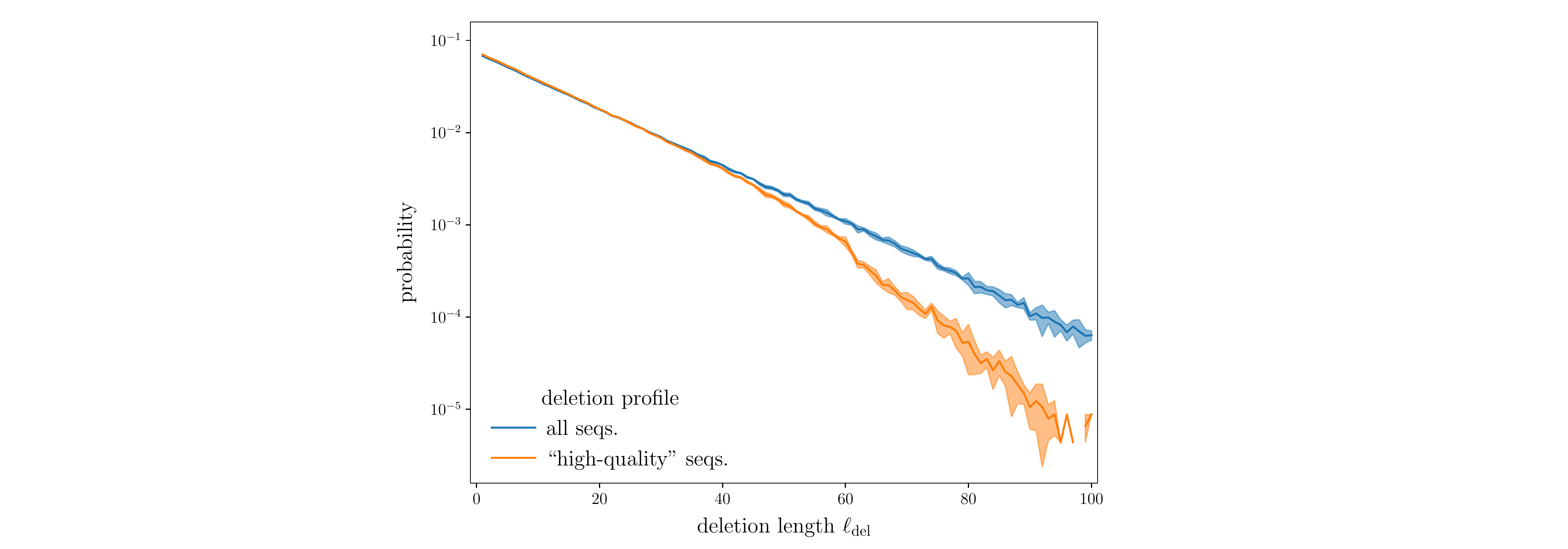}
\caption{\rev Distribution of deletion lengths in 5 synthetic repertoires of $500,000$ sequences each. Synthetic sequences were generated as in Fig.~\ref{fig:synth_data_inference}D, but with a cut-off of 100 base pairs instead of 30 for indel lengths, and with indel-to-point-mutation rates $\beta_{\mathrm{del}}=\beta_{\mathrm{ins}}=0.025$. The blue curve shows the true distribution of deletions lengths in the full dataset, while the orange curve shows the same distribution in sequences that have passed the quality filters described in Methods section. Mean and standard deviation over the 5 subsets shown.}
\label{fig:SI_quality_filter}
\end{center}
\end{figure*}

\begin{figure*}[h]
\begin{center}
\includegraphics[width=\textwidth]{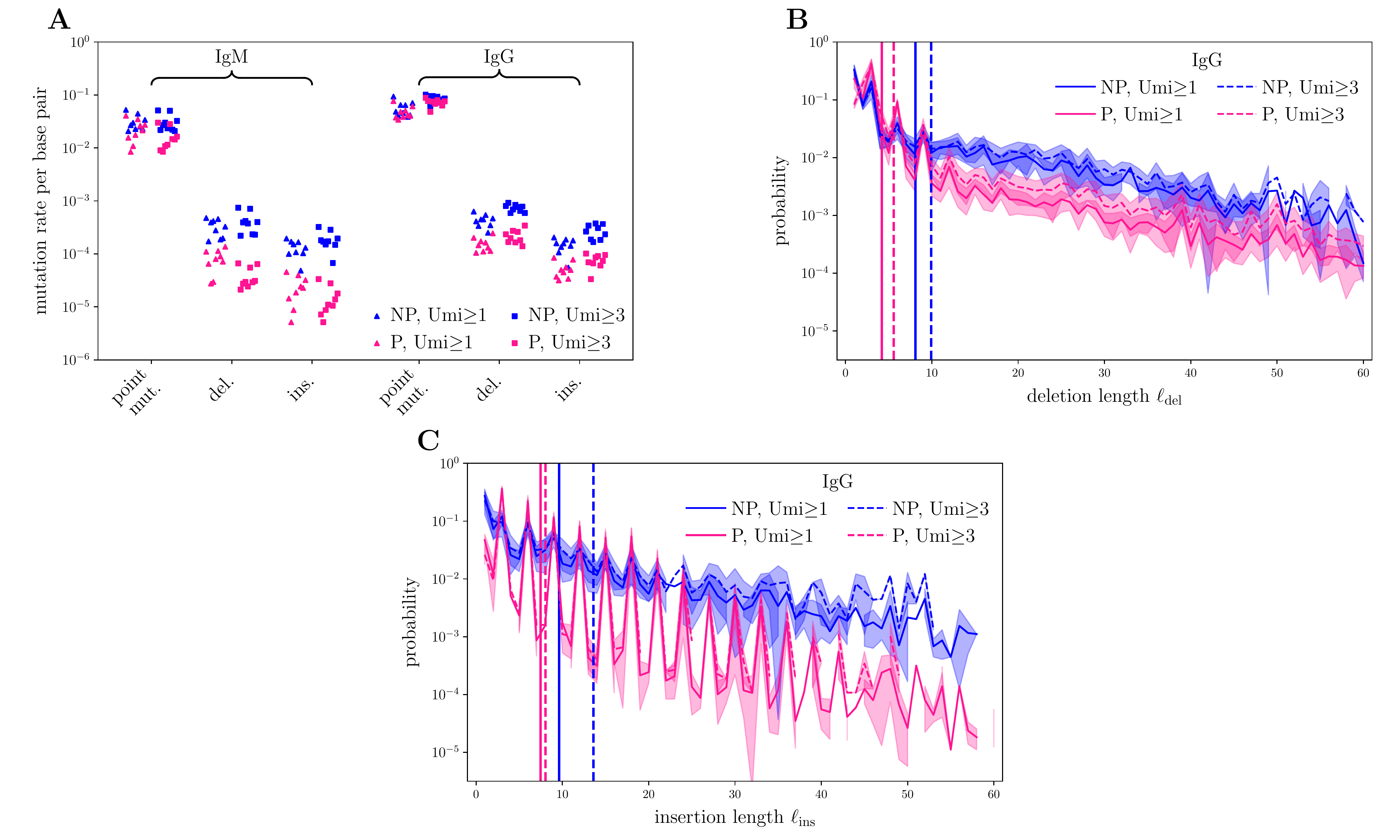}
\caption{\rev Indel statistics from Fig.~\ref{fig:real_data_general_stats}, compared to those obtained from sequences with at least 3 reads per UMI. No systematic error due to sequencing errors (expected to be larger in UMI with low counts) can be detected. \textbf{(A)} Mutation rates per base pair, for both IgM and IgG. \textbf{(B)} Length profiles for deletions in IgG. \textbf{(C)} Length profiles for insertions in IgG. Vertical lines show mean values.}
\label{fig:SI_UMI_check}
\end{center}
\end{figure*}

\end{document}